# Impact of the regularization parameter in the Mean Free Path reconstruction method: Nanoscale heat transport and beyond

**Miguel Ángel Sanchez-Martinez[1], Francesc Alzina[1], Juan Oyarzo[2], Clivia M. Sotomayor Torres[1, 3] and Emigdio Chavez-Angel[1,*]**

[1] Catalan Institute of Nanoscience and Nanotechnology (ICN2), CSIC and The Barcelona Institute of Science and Technology (BIST), Campus UAB, Bellaterra, 08193 Barcelona, Spain; miguel.sanchez.92@gmail.com (MASM), francesc.alzina@icn2.cat (FA), clivia.sotomayor@icn2.cat (CMST), emigdio.chavez@icn2.cat

[2] Instituto de Química, Pontificia Universidad Católica de Valparaíso, Casilla 4059, Valparaíso, Chile.; joyarzo89@gmail.com (JO)

[3] ICREA- Institució Catalana de Recerca i Estudis Avançats, 08010 Barcelona, Spain.

* Correspondence: emigdio.chavez@icn2.cat; Tel.: +34 93 7371617

**Abstract:** The understanding of the mean free path (MFP) distribution of the energy carriers in materials (e.g. electrons, phonons, magnons, etc.) is a key physical insight into their transport properties. In this context, MFP spectroscopy has become an important tool to describe the contribution of carriers with different MFP to the total transport phenomenon. In this work, we revise the MFP reconstruction technique and present a study on the impact of the regularization parameter on the MFP distribution of the energy carriers. By using the L-curve criterion, we calculate the optimal mathematical value of the regularization parameter. The effect of the change from the optimal value in the MFP distribution is analyzed in three case studies of heat transport by phonons. These results demonstrate that the choice of the regularization parameter has a large impact on the physical information obtained from the reconstructed accumulation function, and thus cannot be chosen arbitrarily. The approach can be applied to various transport phenomena at the nanoscale involving carriers of different physical nature and behavior.

**Keywords:** Mean free path reconstruction, Mean free path distribution of phonons, Thermal conductivity distribution

---

## 1. Introduction

In solid-state materials there is a variety of scattering mechanisms for energy carriers involved in different transport phenomena, such as impurities, boundaries, and collisions with other particles/quasi-particles. The average distance that a moving particle (photon, electron, etc.) or quasi particle (phonon, magnon, etc.) travels before being absorbed, attenuated, or scattered is defined as the mean free path (MFP). It is well known that energy carriers that propagate over different distances in a material having different MFP contribute differently to the energy transport. Thus the use of a single-averaged MFP may be inaccurate to describe the system.[1,2]

It is possible to quantitatively describe how energy carriers with a specific MFP contribute to the total transport property by an MFP spectral function or MFP distribution[3], which contains the information of the specific transport property associated with the energy carriers with a certain MFP. By normalizing and integrating this spectral function we obtain the accumulation function, which describes the contribution of carriers with different MFPs below a certain MFP cut off to the total transport property, being very intuitive to identify which MFPs are the most relevant to the transport phenomenon under study by plotting this function.

When studying transport at the nanoscale, boundary scattering becomes important as the characteristic size of the nanostructure approaches the MFP of the carriers involved. From the bulk MFP distribution it is possible to predict how size reduction will affect a transport property in this material given that we know which MFPs are contributing the most. Inversely, it is possible to obtain the bulk MFP distribution from size-dependent experiments, where the critical size of each measurement acts as a MFP cut off due to boundary scattering.

This relation between the transport property at the nanoscale and the bulk MFP distribution is given by an integral transform. A suppression function (SF), which accounts for the specific geometry of the experiment and depends upon the characteristic size of the structures and the MFP of the carriers, connects the bulk MFP distribution and the experimental measurements; acting in the kernel of the integral transformation. Using this integral relation it is possible to recover the MFP distribution from experimental data. This is known as the MFP reconstruction method[4].

This mathematical procedure however is an ill-posed problem with, in principle, infinite solutions. To obtain a physically meaningful result from it, some restrictions must be imposed. These constrains are mainly related to the shape of the mean free path distribution. The distribution is a cumulative distribution function and it is subjected to some restrictions, e.g., the MFP distribution is unlikely to have abrupt steps because it is spread over such a wide range of MFPs. The distribution must thus obey some type of smoothness restriction.[4] Now, the problem becomes a minimization problem, where the solution lies in the best balance between the smoothness of the reconstructed function (solution norm) and its proximity to the experimental data (residual norm). This balance is controlled by the choice of the regularization parameter. The role of this parameter has been widely overlooked in the literature, where the choice of its value has been poorly justified and, in some occasions, it remains unmentioned. In this work, we present a method to obtain the optimal value of this parameter using the *L*-curve criterion[5,6]. We apply it to the thermal conductivity and the phonon-MFP distribution and we present a study of the impact that the choice of its value has in the reconstructed accumulation. We demonstrate that this methodology can be extended to several transport properties involving carriers of different physical nature and behaviour.

## 2. Materials and Methods

To perform the reconstruction of the MFP distribution of the energy carriers, the only input needed is a characteristic SF and a well distributed set of experimental data, i.e., a large amount the experimental measurements spread over the different characteristic sizes of the system. The suppression function strongly depends on the specific geometry of the sample and the experimental configuration. It relates the transport property of the nanostructure $\alpha_{nano}$ and that of the bulk $\alpha_{bulk}$. The suppression has been derived for different experimental geometries from the Boltzmann transport equation [7–9].

The relation between the transport coefficient $\alpha_{nano}(d)$ and the suppression function was originally derived for thermal conductivity [4,10], and more recently has been used to determine the MFP of magnons and the spin diffusion length distribution. [11] This relation can be expressed by means of a cumulative MFP distribution as

$$\alpha = \frac{\alpha_{nano}(d)}{\alpha_{bulk}} = -\int_0^{\infty} F_{acc}(\Lambda_{bulk}) \frac{dS(\chi)}{d\chi} \frac{d\chi}{d\Lambda_{bulk}} \mathrm{d}\Lambda_{bulk} \tag{1}$$

where $S(\chi)$ is the suppression function, $\chi$ is the Knudsen number $\chi = \Lambda_{bulk}/d$, $\Lambda_{bulk}$ is the bulk MFP and $d$ the characteristic length of the sample and $F_{acc}$ is the accumulation function given by

$$F_{acc} = \frac{1}{\alpha_{bulk}} \int_0^{\Lambda_c} \alpha_{\Lambda}(\Lambda_{bulk}) \mathrm{d}\Lambda_{bulk} \tag{2}$$

where $\alpha_{\Lambda}$ is the contribution to the total transport property of carriers with a mean free path $\Lambda$. This function represents the contribution of carriers with MFPs up to an upper limit, $\Lambda_C$, to the total transport property, and is the object that will be recovered by applying the MFP reconstruction technique. From this definition it is easy to see that the accumulated function is subject to some physical restrictions: it cannot take values lower than zero for $\Lambda_c = 0$ or higher than one for $\Lambda_c \to \infty$, and it must be monotonous.[4] We can recognize that **Eq.** (2) is a Fredholm integral equation of the first kind that transforms the accumulation function $F_{acc}(\Lambda_{bulk})$ into $\alpha$ with $K = \frac{dS(\chi)}{d\chi}\frac{d\chi}{d\Lambda_{bulk}}$ acting as a kernel. As the inverse problem of reconstructing the accumulation function $F_{acc}$ is an ill-posed problem with infinite solutions.[10] Minnich demonstrated some restriction can be imposed on $F_{acc}$ to obtain a unique solution.[4] Furthermore, it is reasonable to require the smoothness conditions mentioned before on $F_{acc}$, since it is unlikely to have abrupt behaviour in all its domain. These requirements can be applied through the Tikhonov regularization method, where the criterion to obtain the best solution $F_{acc}$ is to find the following minimum:

$$\min\{\|A \cdot F_{acc} - \alpha_i\|_2^2 + \mu^2 \|\Delta^2 F_{acc}\|_2^2\} \tag{3}$$

where $\alpha_i$ is the normalised $i$-th measurement, $A = K(\chi_{i,j}) \times \beta_{i,j}$ is an $m \times n$ matrix, where $m$ is the number of measurements and $n$ the number of discretization points, $K_{i,j}$ is the value of the kernel at $\chi_{i,j} = \Lambda_{i,j}/d_i$, and $\beta_{i,j}$ the weight of this point for the quadrature. The operators $\| \quad \|_2$ and $\Delta^2$ are the 2-norm and the (n-2)×n trigonal Toeplitz matrix which represent a second order derivative operator, respectively. The first term of Eq. (3) is related to how well our result fits to experimental data (residual norm), while the second term represents the smoothness of the accumulation function (solution norm). The balance between both is controlled by the regularization parameter $\mu$. In other words $\mu$ sets the equilibrium between how good the experimental data is fitted and how smooth is the fitting function. The choice of $\mu$ will thus have a huge impact on the final result of the accumulation function, and a criterion to obtain the optimal value must be established. The selection of the most adequate $\mu$ is still an open question in mathematics. Several heuristic methods are frequently used, such as the Morozov's discrepancy principle, the Quasi-Optimally criterion, the generalized cross validation, *L*-curve criterion, the Gfrerer/Raun method, to name a few.[5,6] Among these methods, the *L*-curve criterion is one of the most popular due to its robustness, convergence speed and efficiency. This method establishes a balance between the size of the discrepancy of the fitting function and the experimental data (residual norm) with the size of the regularized solution (solution norm) for different values of $\mu$. As is shown in Figure 2a, the curve has an L-like shape composed by a steep part where the solutions are dominated by perturbation errors and a flat part where the solution is dominated by regularization errors. The corner represents a compromise between a good fit of the experimental data and the smoothness of the solution. It has been found that the corresponding point lies in the corner of the L-curve in the residual norm-solution norm plane, which can be defined as the point of maximum curvature. The optimal value of $\mu$ can be found by locating the peak position of the curvature as a function of $\mu$.[5,6]

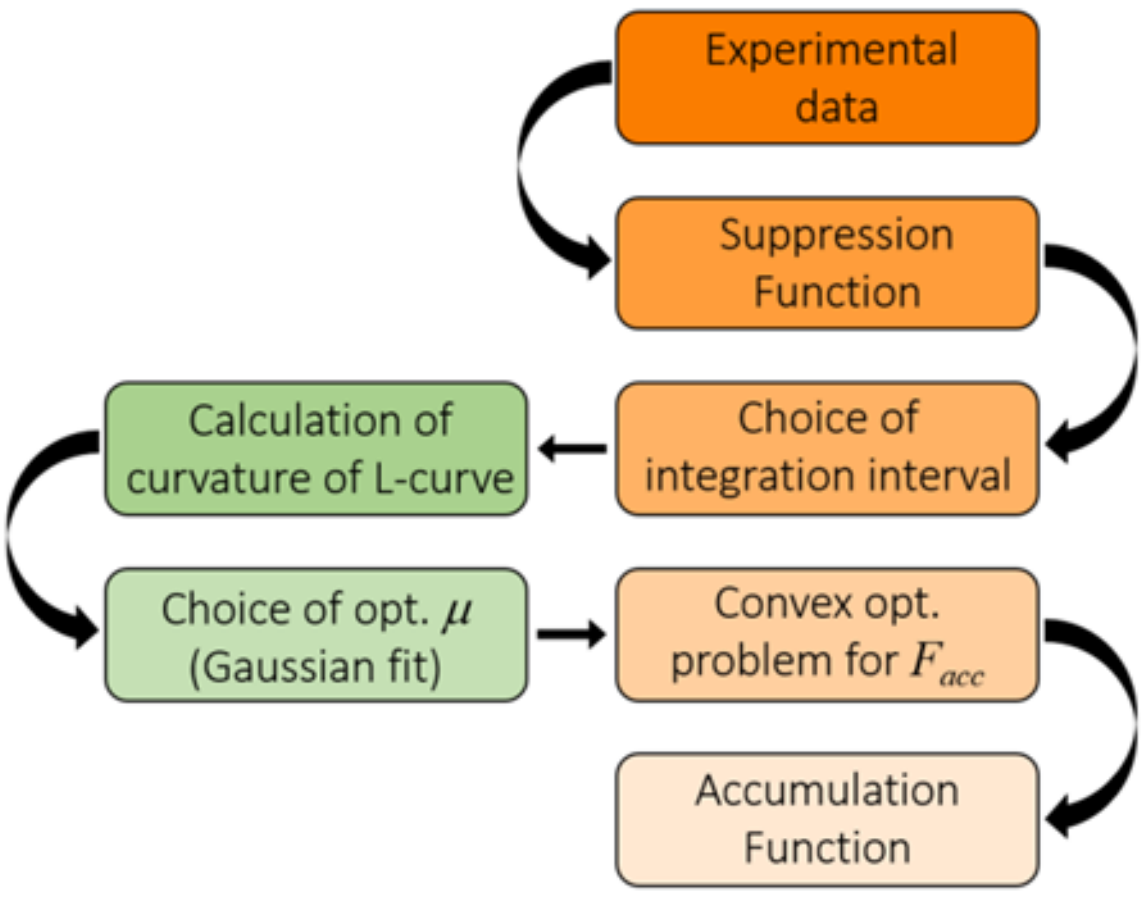

**Figure 1** Flow-chart of the reconstruction method used to obtain the accumulation function from experimental data

The method employed is depicted in Figure 1, and is common to all cases presented here. The experimental data in the case studies shown in this work was reproduced from the digitalization of the images and the corresponding uncertainties. A specific suppression function is selected for each case, depending on the particular geometry and experimental configuration. As explained above, the integral in Eq. (2) is discretized, and an adequate integration interval depending on the span of experimental data is chosen. At this point, instead of introducing an arbitrary value of the regularization parameter, we determine the optimal value via the *L*-curve criterion. We have observed that the distribution of the curvature depending on log $\mu$ follows a Gaussian distribution, thus allowing us to obtain the peak, i.e., the highest curvature (corresponding to the corner of the *L*-curve), using a Gaussian fitting with a reduced number of computational points. With this optimal value of the regularization parameter we can proceed to apply the Tikhonov regularization method and impose the conditions on $F_{acc}$ using a convex optimization package for MATLAB called CVX. [12,13]

## 3. Results and discussions

The MFP reconstruction method, as well as the method here presented for the selection of $\mu$, does not require a priory any physical assumption about the carrier, such as band structure or velocity. In this section, we apply the method to heat transport by phonons in three examples cases where the characteristic size is given by (a) the thickness of the nanostructure in a layered system, (b) a length scale of the measurement technique and (c) a combination of the former two. The extension of the method to magnon-mediated transport phenomena, namely, spin Seebeck effect and spin-Hall torque coefficient is further examined in the supporting information.

### *3.1. Phonons in out-of-plane thermal transport in graphite from molecular dynamic simulations*

Firstly, we will study the MFP reconstruction of phonons in cross-plane thermal transport along the c-axis of graphite. The thermal conductivities ($\kappa$) were obtained from a set of simulations performed by Wei et al.[14] From this numerical experiment, we have recovered the MFP distribution in bulk graphite along the c-axis by using the *L*-curve criterion based MFP reconstruction method. This will illustrate the different steps and details which apply to all of the cases that we present here.

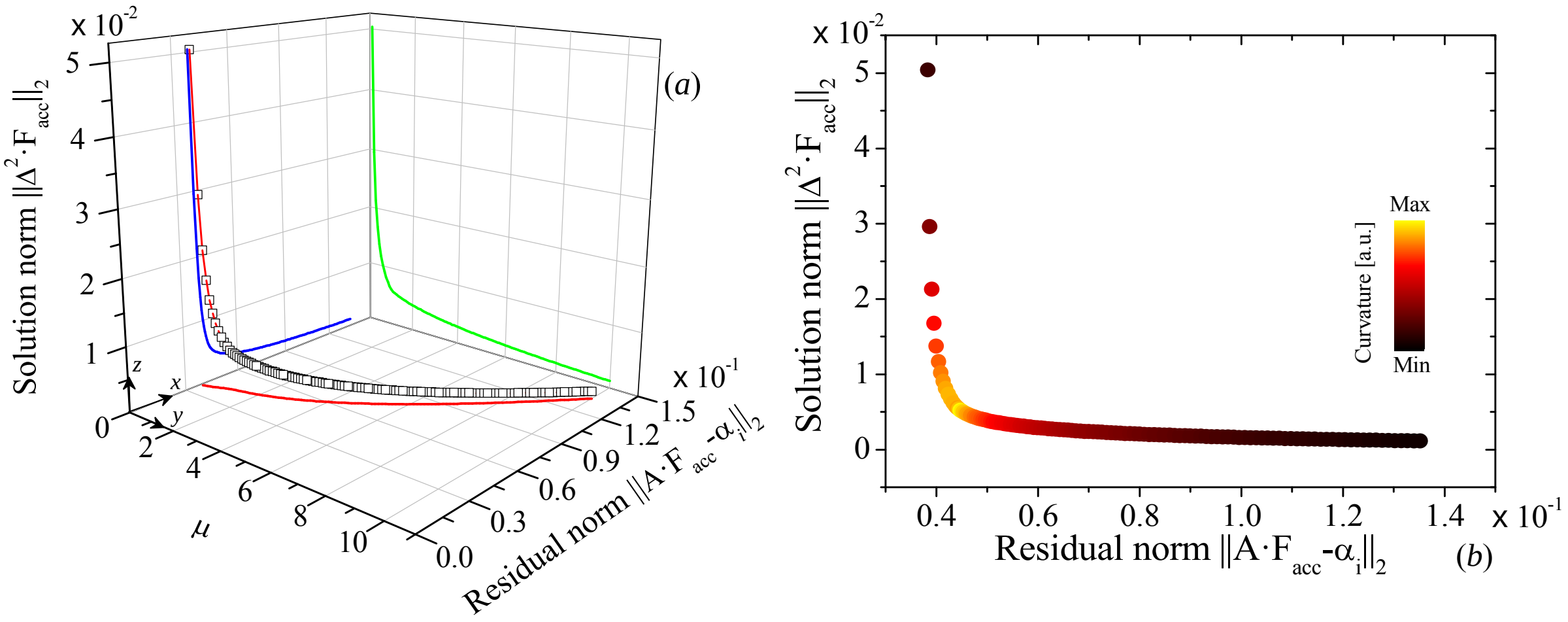

Figure 2 (a) three-dimensional visualization of the relation between the L-Curve (blue) and the different values of μ for a cross-plane Fuchs-Sondheimer suppression function at T= 300 K for graphite simulations. (b) L-curve or *xz* projection of 3D curve, the maximum curvature is displayed as heat-like colour bar.

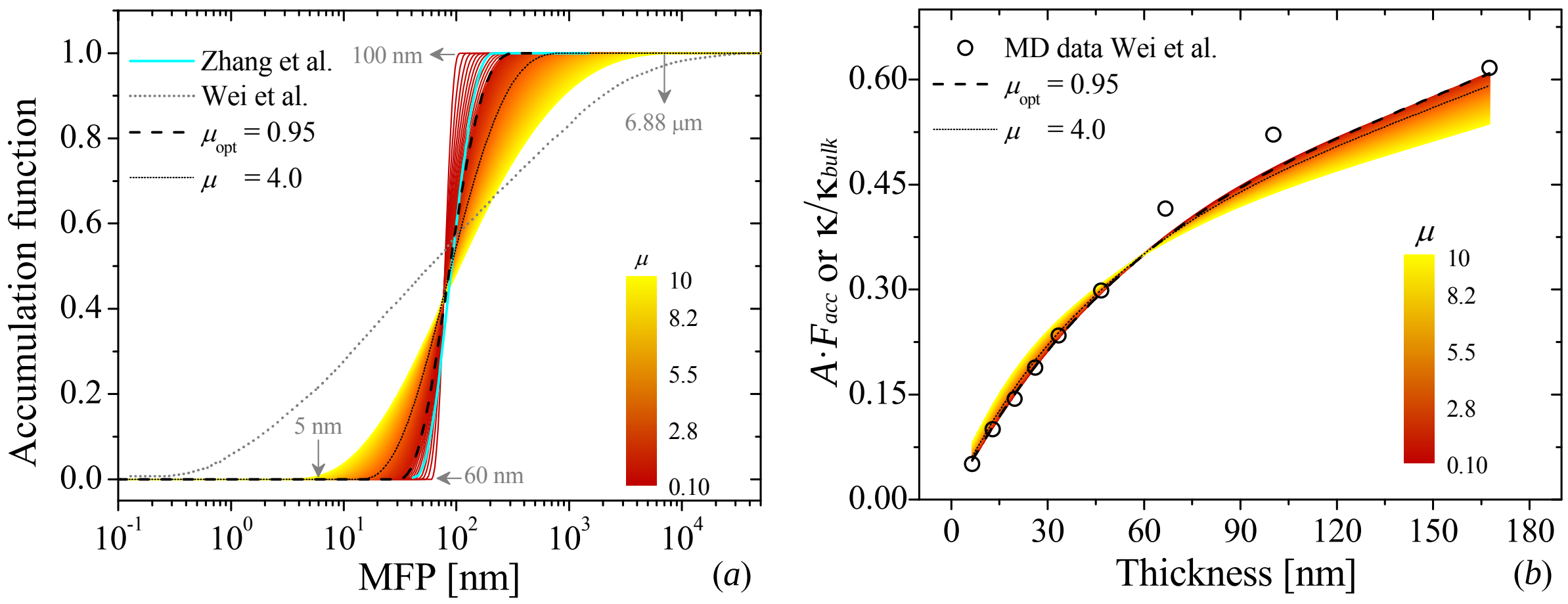

Figure 3 (a) Phonon mean free path distribution of the accumulation function reconstructed for different values of $\mu$ (0.1-10) using the FS cross-plane suppression function and the simulated $\kappa$ made by Wei et al. [14]. The black-dashed and black-dotted lines represent reconstructions obtained by using an optimum (0.95) and large (4.0) μ-value, respectively. The cyan solid and grey dotted line represent the MFP reconstruction obtained by Zhang et al. [15] and Wei et al. [14] (b) Thermal conductivity normalized to the bulk value and/or $A{\cdot}F_{acc}$ product as function of graphite thickness. The open dots represent the simulated thermal conductivity ("experimental data"). Heat-like lines correspond to different $A{\cdot}F_{acc}$ for $\mu$-values in a range of $0.1< \mu <10$.

As shown in Figure 1 after obtaining the "experimental data" the only input needed is the SF. As the thermal transport occurs in the c-direction, i.e., in the cross-plane direction, we will employ the cross-plane Fuchs-Sondheimer SF, described by [16]

$$S(\chi) = 1 - 3\chi\left(\frac{1}{4} - \int_0^1 y^3 \mathrm{e}^{-1/(\chi y)} \mathrm{d}y\right) \tag{4}$$

Equation (4) describes the modification of the phonon transport along the perpendicular direction of a thin film due to the finite size effect. Once fixed the SF, the next step is to find the optimum $\mu$-parameter by minimizing Eq. (3) for different $\mu$-values. The dependence of both the residual and the solution norms on $\mu$ can be visualized as a 3-D curve (see Figure 2a), and its projection on $xz$-plane is the $L$-curve, named after its characteristic shape. The curve was generated using the simulated cross-plane $\mu$ of graphite as function of thickness at T = 300 K. [14] Figure 2b shows the curvature distribution of $\|A \cdot F_{acc} - \alpha_i\|_2$ vs $\|\Delta^2 F_{acc}\|_2$ expressed in heat-like colour map representing the maximum (yellow) and minimum (black) values of the curvature. We can observe that the maximum of the curvature is found right at the corner of the $L$-curve for $\mu \approx 0.95$, which is extracted from the Gaussian distribution of the curvature. In this corner the balance between a smooth function and a good fitting of the experimental data can be found. Figure 3a shows the $F_{acc}$ as a function of the MFP for different $\mu$-values. For this particular example, we can observe that depending on the $\mu$-values the span of MFP can fluctuate from a very wide (5 nm-6.88 μm for $\mu = 10$) to very narrow (60-100 nm for $\mu = 0.1$) distribution. It is also important to notice $\mu$-values between 0.1-4.0 lead to quite good agreement between the simulated $\kappa$ ("experimental data") and $A{\cdot}F_{acc}$ product (see Figure 3b), although these values yield to a completely different span of the MFP. From very narrow (low $\mu$-values) to very wide (high $\mu$-values) distribution. This difference in the MFP distribution is a direct consequence of the weight given to each component in the Eq. (3). The low $\mu$-values give a large weight to the residual norm and, as a consequence, it leads to an overfitting of the reconstructed function. While large $\mu$-values give higher importance to the solution norm leading to and over smoothing of the reconstructed function.

Figure 3a also shows the MFP reconstruction generated by Zhang et al. [15] (cyan solid line) and by Wei et al. (grey dotted line). The first reconstruction was carried out using the measured $\kappa$ of graphite as function of the thickness and the suppression function given by equation (4). We can see that for the optimal value of $\mu = 0.95$, the MFP of the carriers spans 36 nm < $\Lambda$ < 270 nm, in similar range compared with the reconstructed values obtained by Zhang et al. 40 nm < $\Lambda$ < 210 nm. [15] The second reconstruction was calculated using the simulated $\kappa$ and a quasi-ballistic model to describe the suppression function (Equations 3 and 4 of ref. [14]). The drastic change of the MFP-distribution is due to that the reconstruction method is an ill-posed problem, then, a drastic change of SF can have a strong impact on its distribution. The desired $F_{acc}$ will be strongly affected by the "shape" of the selected SF. Therefore the choice of a different SF will lead to a different MFP distribution.[11]

*3.2. In-plane thermal transport in 400nm thick Si membrane*

The second case of study is the in-plane thermal transport in a 400 nm Si membrane at room temperature probed by the Thermal Transient Grating (TTG) technique, obtained from Johnson et al.[17] In the previous example the characteristic length of the system was the thickness of the graphite sheets. Here the Si membrane has fixed thickness $d = 400$ nm and the variable length scale is the period of the thermal grating $L$. Since the measurement of in-plane thermal conductivity in the membrane will correspond to phonons with MFPs lower than L in each measurement, the grating period becomes the equivalent of the maximum MFP.

In this case we have to consider two different effects: (i) the impact of the finite size of the membrane and the boundary scattering of the effective in-plane MFP and (ii) the crossover from non-diffusive to diffusive phonon transport given by the period of the thermal grating. Then, a combination of two suppression functions have to be introduced that takes into account both effects. [8]

In the first place, we define an effective MFP ($\Lambda'$) to take into account the effect of the boundary scattering due to thickness of the membrane as follow:

$$\Lambda' = \Lambda S_2(d/\Lambda) \tag{5}$$

where $\Lambda$ is the bulk MFP, d is the thickness of the membrane and $S_2$ is Fuchs-Sondheimer SF for in-plane thermal transport given by [10]

$$S_2(d/\Lambda) = 1 - \frac{3}{8}\frac{d}{\Lambda} + \frac{3}{2}\frac{d}{\Lambda}\int_1^{\infty}(y^{-3} - y^{-5})\mathrm{e}^{-y\Lambda/d}\mathrm{d}y \tag{6}$$

With this effective MFP we proceed to perform the reconstruction using the suppression function for the specific geometry of the experiment given by [17]

$$S_1(q\Lambda') = \frac{3}{q^2\Lambda'^2}\left(1 - \frac{\arctan(q\Lambda')}{q\Lambda'}\right) \tag{7}$$

where $q = 2\Lambda/L$ and $L$ is the grating wavevector. If we define $\zeta = q\Lambda'$, the kernel for the reconstruction is given by

$$\frac{dS(\zeta)}{d\zeta}\frac{d\zeta}{d\Lambda'}\frac{d\Lambda}{d\Lambda'} = \frac{2\pi}{L}\frac{3}{\zeta^3}\left(\frac{3\arctan(\zeta)}{\zeta} - \frac{3+2\zeta^2}{1+\zeta^2}\right)\left(1 + \frac{3}{2}\int_1^\infty dt(t^{-3} - t^{-5})e^{\frac{-t\Lambda}{d}}(1-t)\right) \quad (8)$$

The $S_1$ function is unity in the limit $q\Lambda' \ll 1$, in the diffusive limit, and goes like $(q\Lambda')^{-2}$ for $q\Lambda' \gg 1$, in the ballistic regime, thus describing the transition between both regimes necessary to adequately interpret the measured quantities in the experiment.[17]

Similarly to the previous case, we apply a Gaussian fitting procedure to obtain a value $\mu_{opt} = 1.05$ at room temperature. In Figure 4 we can observe the effect of the different values of $\mu$. Note that, in this case, the reduction of $\mu$ affects mainly the smoothness of the reconstructed function with large changes on the concavity and convexity of the accumulation function. However, there was not a significant change on the $A{\cdot}F_{acc}$ product for small $\mu$-values. The oscillations observed for low $\mu$-values are a direct consequence of the overfitting of the reconstructed function, due to the large weight imposed to the minimization problem. The increase of the regularization parameter beyond the optimal value results in an increase of the span of the MFP of the carriers, as shown in Figure 4a, and a poor agreement with the experimental data, as can be seen in Figure 4b.

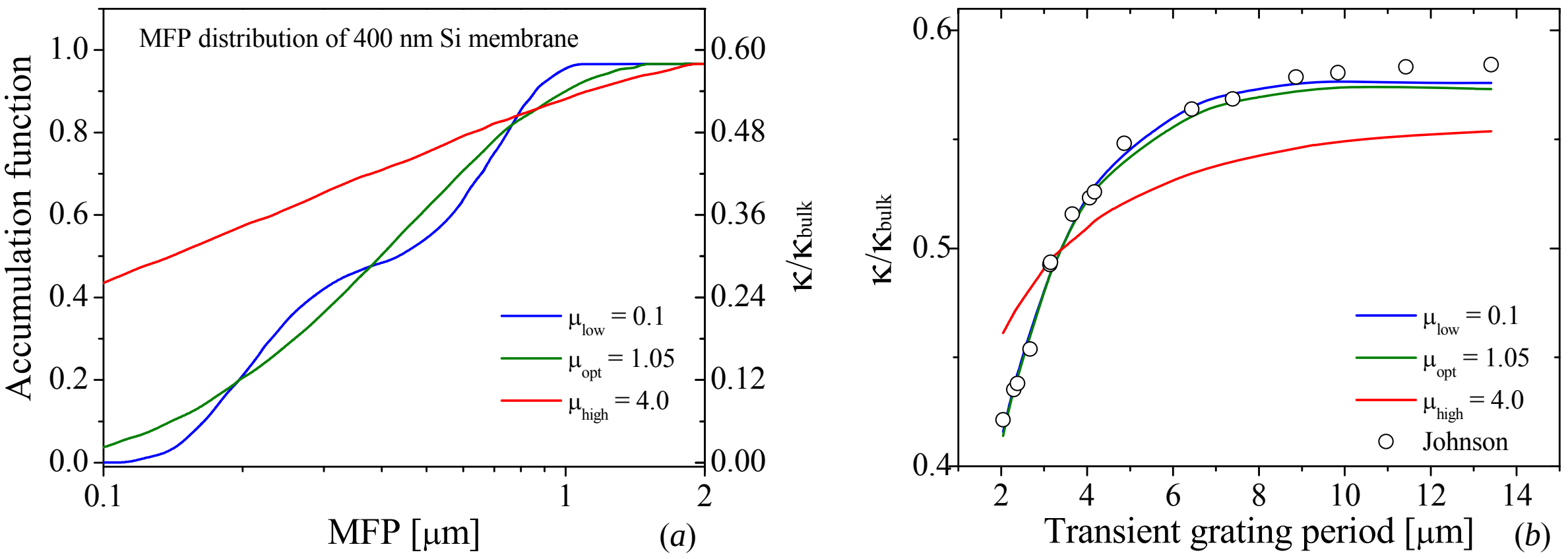

Figure 4 (a) Phonon mean free path distribution of the accumulation function for a 400 nm Si film reconstructed for $\mu_{opt}$ (green line). The blue and red lines are the result obtained using a low and high value of $\mu$, respectively. (b) Thermal conductivity of 400 nm Si film corresponding to the different accumulation functions (red, green, and blue lines) for different transient grating periods in the experimental technique. [17]

### *3.3. In-Plane Thermal transport in Si: Reconstruction by changing the thickness of the membrane*

The dependence of the reconstructed accumulation function on the regularization parameter relies on the $L$-curve and the dependence of the residual norm $||A \cdot F_{acc} - \alpha_i||_2$ and solution norm $||\Delta^2 F_{acc}||_2$ on $\mu$. The following case illustrates an example of how the particular shapes of these different curves can strongly affect the reconstruction.

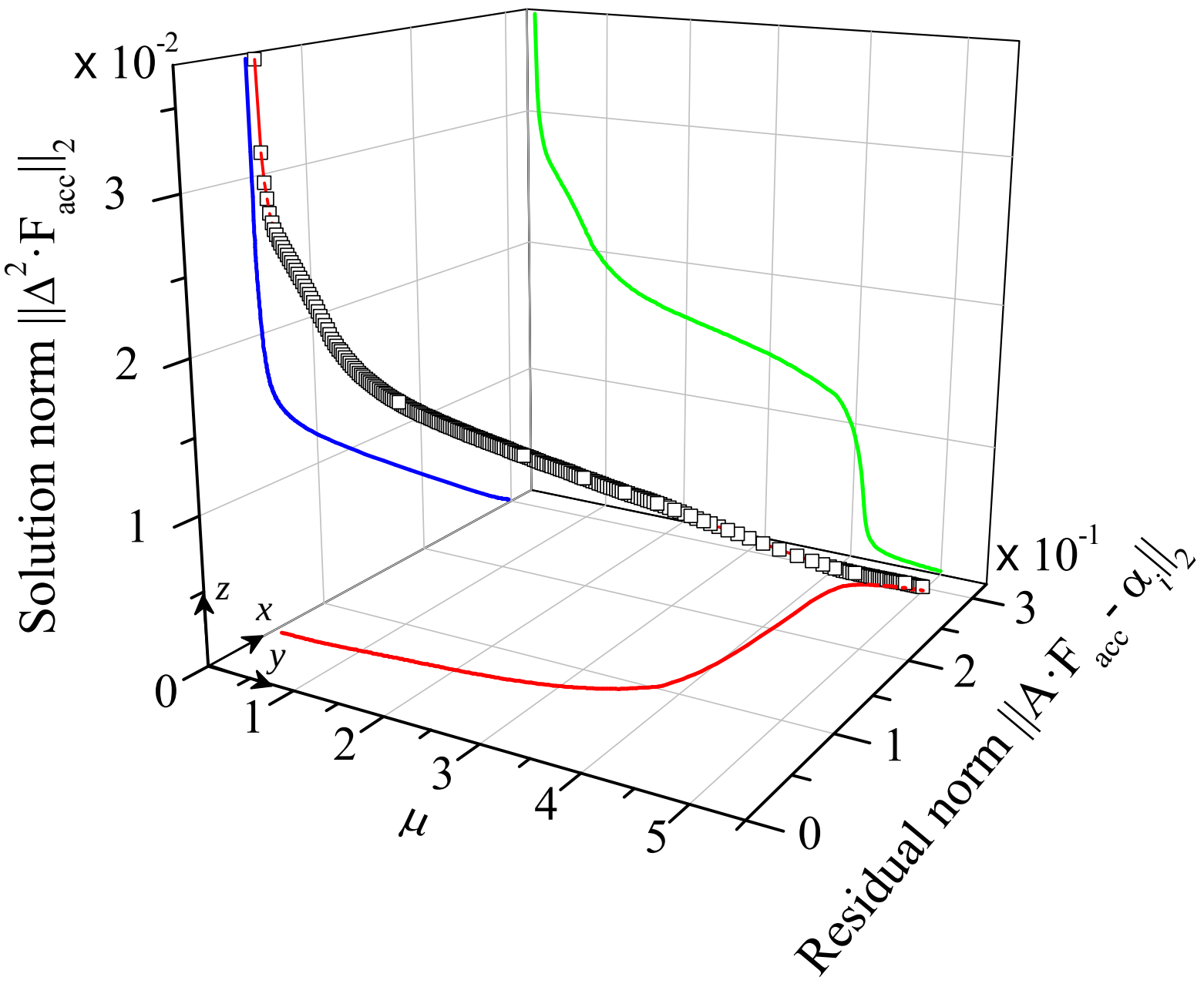

Figure 5 three-dimensional visualization of the relation between the L-Curve (blue) and the different values of $\mu$ for an in-plane Fuchs-Sondheimer suppression function at T= 300 K for silicon measurements.

For this example, we used the thickness dependence of the in-plane $\kappa$ measured by Cuffe et al. [10] The experiment consisted in the measurement of the $\kappa$ of free standing Si membranes ranging from 15- 1518 nm using the transient thermal grating technique in the diffusive regime. As the thermal transport occurs along the films, the Fuchs-Sondheimer for in-plane thermal transport (see Eq. (6)) was selected as SF.

As we can see in Figure 5, the three-dimensional (3D) visualization of the minimization problem is manifestly different to that presented for graphite in Figure 2a. For graphite, we observed very smooth behaviour of each of the projections (*xy*, *xz* and *yz* planes) in the 3D curve. This leads a continuous variation of MFP from a narrow to a wide distribution as $\mu$ increases. On the other hand, Figure 5 shows singularities and abrupt jumps of the different projections. For the case of low $\mu$-values, this lead to large oscillations in the MFP distribution function as it is shown in blue solid line in Figure 6a. While for large $\mu$-values, we observe a linear distribution of the reconstructed curve as function of the logarithm of MFP (see red solid line in Figure 6a). For intermediate $\mu$-values (0.8-3), we can observe no major differences neither in the smoothness nor in the MFP span of the accumulated function. Similarly, we did not observe huge changes in the fitted function as it is display in Figure 6b.

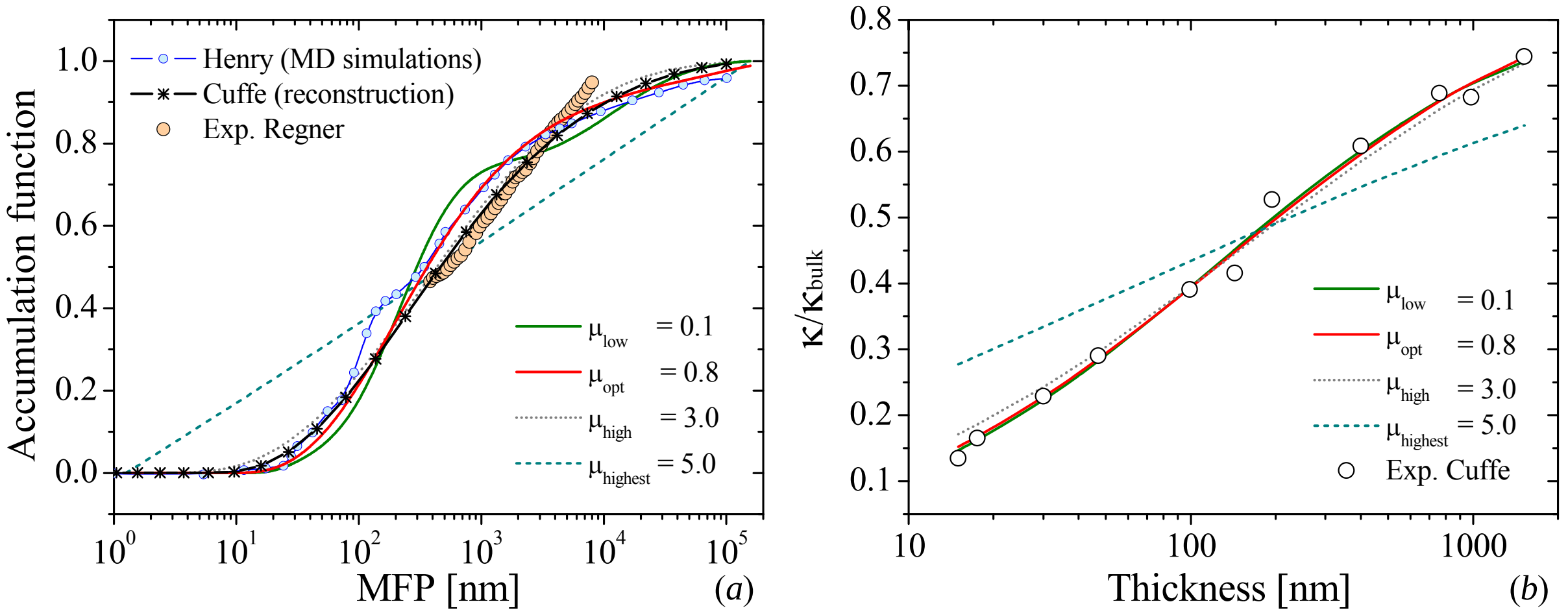

Figure 6 (a) Phonon mean free path distribution of bulk Silicon reconstructed from the thickness dependence of thermal conductivity. The dark green, red, dotted grey and dotted dark cyan lines represent reconstructions for different $\mu$-values. The black stared and blue dotted lines represent the MFP reconstruction and molecular dynamic calculation performed by Cuffe et al. [10] and Henry et al. [18], respectively. The orange-solid dots represent the experimental phonon-MFP distribution of bulk silicon measured by Regner et al.[19] (b) Thermal conductivity corresponding to the different accumulation functions for different samples with different thickness[10].

Figure 6*a* also shows the MFP reconstruction generated by Cuffe et al. [10]. This reconstruction was carried out using the same procedure than the showed in this section including the same suppression function and the minimization procedure. However, no information is given regarding the regularization parameter or the procedure to estimate it. Despite of that, we can observe that MFP distribution estimated for the optimal value of $\mu = 0.8$ and the one obtained by Cuffe et al. spans in similar range with MFP ranging from 20 nm < $\Lambda$ < 100 µm. Slight deviations can be observed for MFP around $\Lambda = 1$ µm, which are expected from small differences in the regularization parameter and the numerical calculation of the minimization problem. For comparison, the calculated and measured MFP distribution of bulk Si are also included. It is very remarkable the good agreement between our reconstruction distribution and the molecular dynamics simulations by Henry et al. [18] and the measurements performed by Regner et al. [19] In the latter case, it is important to keep in mind that the experiment is completely different from the one presented here.

The L-curve criterion is efficient to obtain the most adequate regularization parameter, but in this case the reconstructed accumulation function is very robust against changes in $\mu$ due to the particular flatness of the residual and solution norm depending on the values of $\mu$. It is also important to remark that the estimation of the optimum $\mu$-value has to be carried out for each set of measurements. As we showed in the supporting information, for the case of temperature dependence measurements of the Spin Seebeck coefficient, the optimal $\mu$-value varies among the measurements (See Figure S2). This effect comes from the differences in the spread of the experimental data in each temperature regime and the importance of carriers with longer MFPs as the temperature decreases.[11]

## 4. Conclusions

We have demonstrated that the choice of the regularization parameter $\mu$ has a large impact on the physical information obtained from the reconstructed accumulation function, and thus cannot be chosen arbitrarily. On one hand, small $\mu$-values lead to a very narrow MFP distribution but very good agreement between the fitting function and the experimental data. However, it also introduces artefacts (oscillations) to the reconstructed MFP distributions. These oscillations are consequence of an overfitting of the reconstructed function due to the low weight given to the solution norm. On the other hand, large $\mu$-values lead to a wider MFP distribution but worse agreement between the fitting function and the experimental data. This is a direct consequence of the larger weight to the solution norm and, as consequence, an over smoothing of the fitted function. It is also important to notice that the estimation of the optimum $\mu$-value has to be carried out for each set of measurements. It cannot be assumed the same value for temperature dependence measurements as we show it in the supporting information.

The method presented here is not limited only to phonon transport and it can be applied to many different cases. The only input needed to reconstruct the mean free path distribution of the carriers is a well-known suppression function and a well distributed set of experimental data points. Regardless of the carrier physical behaviour or the span of MFP, we have proven the method to be applicable, and we have seen that the robustness of the reconstruction against deviations from the estimated value of $\mu$ will depend on the particular variation of the solution norm and reduced norm with $\mu$.

**Supplementary Materials:** The following are available online at www.mdpi.com/xxx/s1, Figure S1: (a) Magnon mean free path distribution of the accumulation function for spin Seebeck effect experiments. (b) Normalized Longitudinal Spin-Seebeck coefficient for different thickness of the YIG. Figure S2: Optimal values of $\mu$ for the Magnon-MFP reconstruction. Figure S3: (a) Spin diffusion length distribution of the accumulation function reconstructed for spin-Hall torque experiments. The blue and red lines are the result obtained using a low and high value of $\mu$, respectively. The green line represent the MFP distribution reconstructed using $\mu_{opt}$. (b) Normalized Longitudinal spin-Hall torque coefficient for different thickness of the Pt filmvalues of μ for the Magnon-MFP reconstruction at different temperatures for LSSE experiments.

**Author Contributions:** MASM, FA, ECA wrote the article. MASM, JO and ECA designed the different numerical programs. CMST and FA supervised the work and discussed the results. All authors discussed the results and commented on the manuscript.

**Funding:** This research was funded by the EU FP7 project QUANTIHEAT (Grant No 604668) and the Spanish MINECO-Feder project PHENTOM (FIS2015-70862-P). M.A.S.M. acknowledges support from SO-FPI fellowship BES 2015-075920. ICN2 acknowledges support from Severo Ochoa Program (MINECO, Grant SEV-2017-0706) and funding from the CERCA Programme/ Generalitat de Catalunya.

**Acknowledgments:** We thank Dr. G. Whitworth for a critical reading of the manuscript.

**Conflicts of Interest:** The authors declare no conflict of interest.

# Impact of the regularization parameter in the Mean Free Path reconstruction method: Nanoscale heat transport and beyond

M.A. Sánchez-Martínez,[1] F. Alzina, [1] J. Oyarzo, [2] C.M. Sotomayor-Torres,[1, 3] and E. Chavez-Angel[*1]

[1)] Catalan Institute of Nanoscience and Nanotechnology (ICN2), CSIC and The Barcelona Institute of Science and Technology (BIST), Campus UAB, Bellaterra, 08193 Barcelona, Spain.

[2)] Instituto de Química, Pontificia Universidad Católica de Valparaíso, Casilla 4059, Valparaíso, Chile.

[3)] ICREA- Institució Catalana de Recerca i Estudis Avançats, 08010 Barcelona, Spain.

*Corresponding author: emigdio.chavez@icn2.cat

## Magnon-mediated longitudinal spin-Seebeck effect in YIG films

This data was obtained by Guo *et al* measuring the thickness dependence of the longitudinal-Spin Seebeck effect (LSSE) in YIG films.[1] The suppression function used in the reconstruction was the cross-plane Fuchs-Sondheimer model presented in Eq. (4). In **Figure S1*a*** it is easy to see that the different values of $\mu$ have an impact mainly in the accumulation function, being the case of the lower value of $\mu$ that affects the most both the accumulation function and the recovered LSSE, as shown in Figure S1***b***. The reconstructed function for the optimal value $\mu_{opt}$ = 0.8856 at T=250 K is shown in Fig. 10. The optimal value of the reconstruction parameter was calculated for all the temperatures measured, obtaining a different value for each of them (see **Figure S2**).

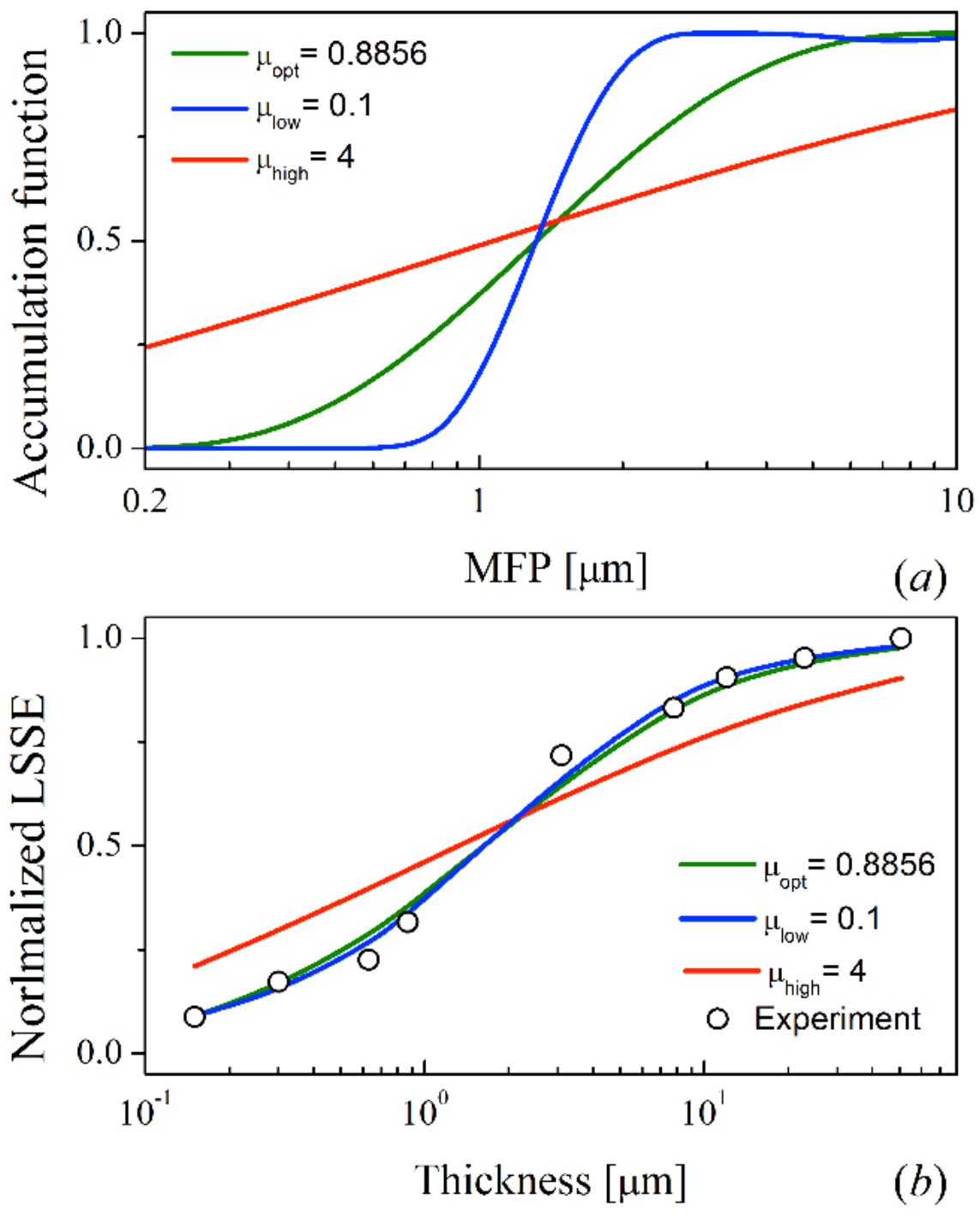

Figure S1 *(a) Magnon mean free path distribution of the accumulation function reconstructed for m$_{opt}$ (green line) for spin Seebeck effect experiments. The blue and red lines are the result obtained using a low and high value of m, respectively. (b) Normalized Longitudinal Spin-Seebeck coefficient for different thickness of the YIG sample.* [1]

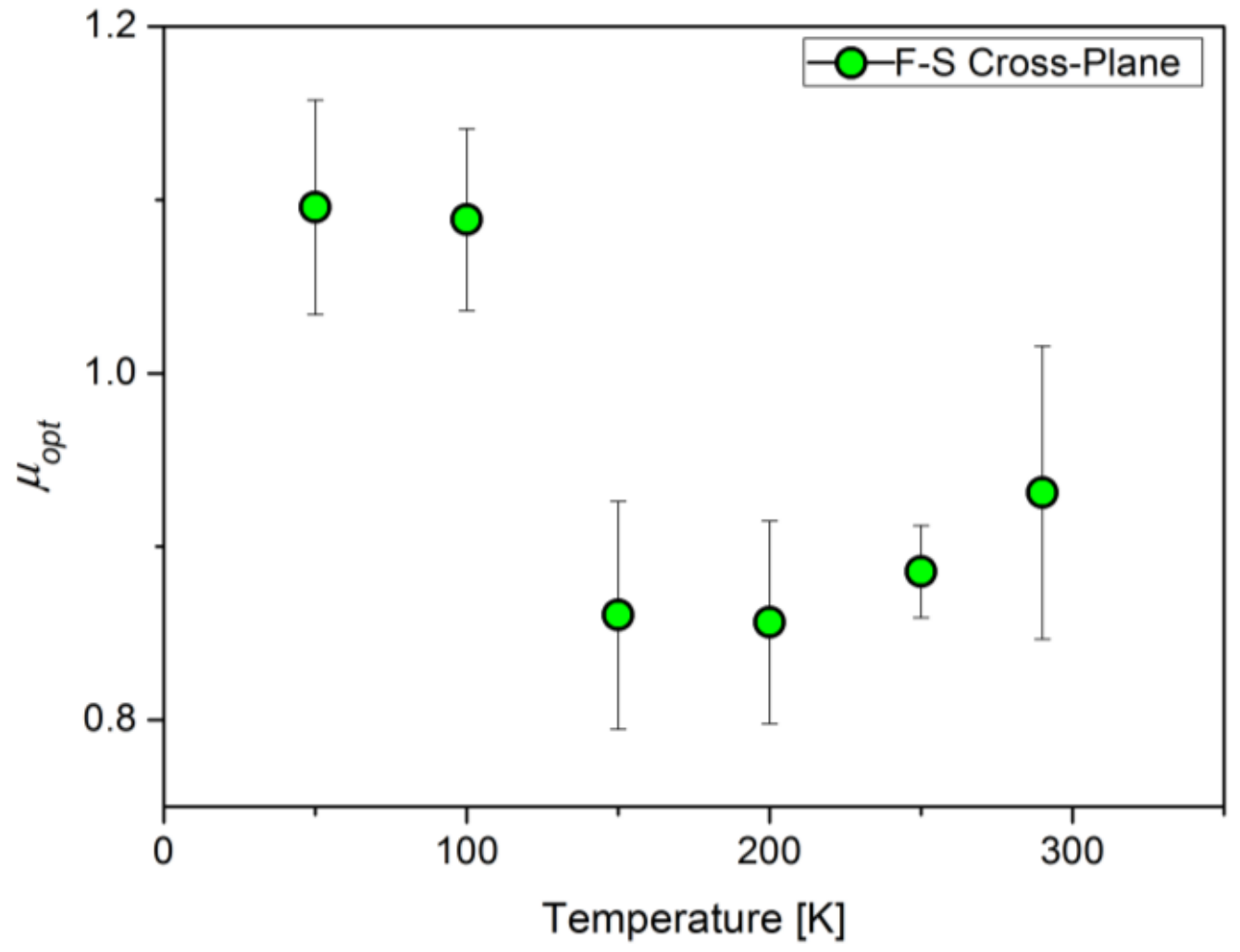

Figure S2 *Optimal values of μ for the Magnon-MFP reconstruction at different temperatures for LSSE experiments.*

## Spin diffusion length in Pt films

This case is an example of the application of this technique to a completely different transport phenomena, namely, the spin diffusion length. The experiment consisted of measuring the thickness dependence of the spin-Hall torque coefficient, $\xi$, in platinum.[2]
In this case, the suppression function is derived from the drift-diffusion model [3], and given by [4]

$$S(\chi) = 1 - \frac{1}{\sinh(1/\chi)} \tag{S1}$$

This case differs from the previous phenomena, having a narrow span of the MFP distribution. For the optimal value of $\mu$, the accumulation function's range goes from around 0.7 to 2 nm, and for $\mu$ = 0.1 it shrinks to be from 1 to 2 nm (see Figure S3). This indicates that the method is very sensitive even for very narrow MFP distributions, which allows it to be applied to a wide range of transport phenomena.

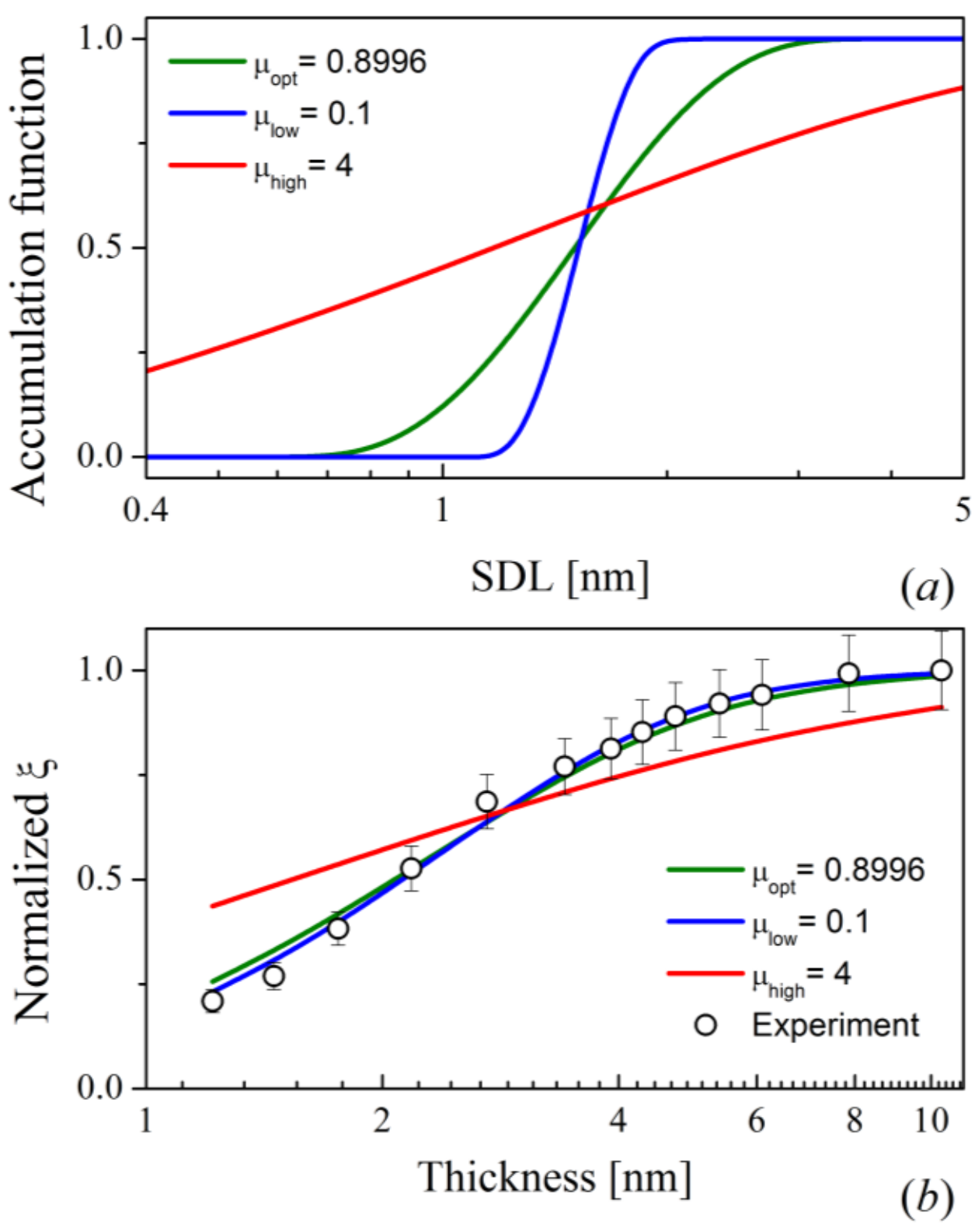

Figure S3 *Spin diffusion length distribution of the accumulation function reconstructed for spin-Hall torque experiments. The blue and red lines are the result obtained using a low and high value of $\mu$, respectively. The green line represent the MFP distribution reconstructed using $\mu_{opt}$. (b) Normalized Longitudinal spin-Hall torque coefficient for different thickness of the Pt filmvalues of $\mu$ for the Magnon-MFP reconstruction at different temperatures for LSSE experiments.*